\documentclass[openacc]{rstransa}
\usepackage{amssymb, amsmath, amsthm,mathrsfs}
\usepackage{mathtools,bm,color}
\usepackage{wrapfig, tikz,url,graphics, float, ulem}
\graphicspath{{fig/}}
\def \tb {t_{\rm b}}
\def \fzero {f_{\rm 0}}

\def \mA {\mathbb{A}}
\def \mZ {\mathbb{Z}}
\def \ab {\alpha\beta}
\def \ac {\alpha\gamma}
\def \cb {\gamma\beta}

\def \XX {{\bm X}}
\def \VV {{\bm V}}

\def \del2z {\partial^{2}_{z}}

\def \rhog {\rho_{\rm g}}
\def \rhod {\rho_{\rm d}}

\def \uu {{\bm u}}

\def \LI {{\mathcal L}_{\rm I}}
\def \tL {T_{\mathcal L}}

\def \curl {{\bm \nabla} \times}
\def \dive {{\bm \nabla}\cdot}

\newcommand{\eq}[1]{(\ref{#1})}

\newcommand{\Eq}[1]{Eq.~(\ref{#1})}

\newcommand{\bra}[1]{\left\langle #1\right\rangle}
\newcommand{\ddt}[1]{\frac{d#1}{dt}}
\newcommand{\Tr}[1]{{\mathcal Tr}\left[#1\right]}

{}

\def \leta {\ell_{\eta}}

\def \Rey  {\mbox{Re}}
\def \Rel  {\mbox{Re}_{\lambda}}

\def \St  {\mbox{St}}
\def \Ku  {\mbox{Ku}}

\def \taup {\tau_{\rm p}}

\def \teta  {\tau_{\eta}}

\def \kf  {k_{\rm f}}

\def \urms  {u_{\rm rms}}
\def \orms {\omega_{\rm rms}}

\def \curl {{\bm \nabla}\times}

\def \kf  {k_{\rm f}}

\def \urms {u_{\rm rms}}

\def \Zth {Z_{\rm th}}

\newcommand{\fig}[1]{Fig.~\ref{#1}}
\newcommand{\subfig}[2]{Fig.~\ref{#1}(#2)}

\def\drawing #1 #2 #3 {
\begin{center}
\setlength{\unitlength}{1mm}
\begin{picture}(#1,#2)(0,0)
\put(0,0){\framebox(#1,#2){#3}}
\end{picture}
\end{center} }
\def \rhog {\rho_{\rm g}}

\newcommand{\beq}{\begin{equation}}
\newcommand{\eeq}{\end{equation}}

\begin{document}
\title{Rate of formation of caustics in heavy particles advected by turbulence }
\author{
  Akshay  Bhatnagar$^1$,
  Vikash Pandey$^2$,
  Prasad Perlekar$^2$, and
  Dhrubaditya Mitra$^4$ }
\address{$^1$ NORDITA, Royal Institute of Technology and Stockholm University,  Stockholm. \\
$^2$ TIFR Centre for Interdisciplinary Sciences, Hyderabad.}
\subject{}
\corres{Dhrubaditya Mitra\\
  email{ dhruba.mitra@gmail.com} }
\date{\today}
\begin{abstract}
  The rate of collision and the relative velocities of the colliding particles
  in  turbulent flows is a crucial part of 
  several natural phenomena, e.g.,
  rain formation in warm clouds and planetesimal formation in a protoplanetary
  disks.
  The particles are often modeled as passive, but heavy and inertial.
  Within this model, large relative velocities emerge due to formation of
  singularities (caustics) of in the 
  gradient matrix of the velocities of the particles. 
  Using extensive direct numerical simulations of heavy particles 
  in both two (direct and inverse cascade) and three dimensional turbulent
  flows we calculate the
  rate of formation of caustics, $J$ as a function of the Stokes number ($\St$).
  The best approximation to our data is $J \sim \exp(-C/\St)$,
  in the limit  $\St \to 0 $
  where $C$ is a non-universal constant. 
\end{abstract}

\begin{fmtext}
  \section{Introduction}
Turbulent flows in nature often have small particles embedded in them.
Two canonical examples are gas flows in proto-planetary disks with
small dust particles~\cite{Arm10} and air flows in a cloud with small
water droplets~\cite{Pruppacher2010microphysics}. 
The first one is  a useful 
model to understand the formation of 
planetesimals --
small kilometer size objects that themselves collide and merge to form
planets. The second one controls the physics of rain formation in warm clouds.
In both of these cases a crucial problem is to understand the growth of few
large objects from a lot of small ones. Let us consider the second case.

\end{fmtext}
\maketitle
Very small water droplets form by condensation in a super-saturated
environment in the cloud. If only condensation and evaporation determines
the evolution of the size of the droplets then, it can be estimated that,
it would take unnaturally long for raindrops to form in a
cloud~\cite{Pruppacher2010microphysics}.  
Clearly, the droplets can also collide with
each other and consequently merge or bounce off. 
The collision between droplets is determined by their relative velocities
at small relative distances. 
If the velocity field of the droplets is smooth everywhere then the
relative velocities between droplets go to zero as their
relative distances go to zero.
In this case, both the frequency of collisions
and collision velocities remains small~\cite{saf+tur56} and the estimated time
to form raindrops is still unnaturally long.
One way out of this conundrum is to consider the possibility that the
velocity field of the droplets does not remain smooth but develop singularities
-- such that the relative velocity between two infinitesimally close droplets
remains finite.

\subsection{Model}
In the simplest case -- the droplets are small and much heavier than the gas --
the velocity of a single droplet in a flow satisfies the
following equations:
\begin{subequations}
\begin{align}
\ddt{\XX(t)} &= \VV(t), \label{eq:dxdt}\\
\ddt{\VV(t)} &= \frac{1}{\taup}\left[\uu(\XX,t) -\VV  \right] , \label{eq:dvdt} 
\end{align}
\label{eq:HIP}
\end{subequations}
where $\XX$ is the position, $\VV$ is the velocity of the droplet, $\uu$ 
is the velocity of the gas at a point $\XX$, and
$\taup \equiv (2\rhod a^2)/(9 \rhog \nu)$ is the relaxation time of the droplet.
Here $a$ is the radius of the droplet, $\nu$ is the kinematic viscosity of the gas,
and $\rhog$ ($\rhod$) is the gas (droplet) density.
We nondimensionalize $\taup$ to introduce the Stokes number:
$\St \equiv \taup/\tau$, where $\tau$ is a characteristic time scale of
turbulence.
A small enough dust grain in a protoplanetary disk also obeys the
same equations. To keep our discussions general, in the rest of this paper,
we shall use the word ``heavy inertial particle'' to mean a water droplet
or a dust grain small enough that their motion obeys \eq{eq:HIP}. 

In the Lagrangian frame of this heavy inertial particle the equation of evolution of
the gradient of its velocity matrix, $\mZ$, with components 
$Z_{\ab} \equiv \partial_{\beta} V_{\alpha}$ ($\alpha,\beta = 1,\ldots, d$ in $d$ dimensions), 
is given by
\begin{equation}
\ddt{Z_{\ab}} + Z_{\ac}Z_{\cb} +  \frac{1}{\taup}  Z_{\ab} =   \frac{1}{\taup} A_{\ab}  \/. 
\label{eq:dA}
\end{equation}
Where $A_{\ab} \equiv \partial_{\beta} u_{\alpha}$ are the components of the  
velocity-gradient matrix of the flow, $\mA$, and repeated indices
are summed.
This equation contains the possibility that elements of
$\mZ$ can become infinitely large in finite time.
To see this first consider the same equation in one dimension. 
Now both the particle velocity-gradient and the fluid velocity-gradients are scalars and 
\eq{eq:dA} simplifies to
\begin{equation}
  \ddt{Z} + Z^2 + \frac{1}{\taup} Z = \frac{1}{\taup} A  \/,
  \label{eq:dA1d}
\end{equation}
where $Z\equiv \partial_x V$ and $A\equiv \partial_x u$.
If we ignore the terms with coefficients equal to $1/\tau_p$ in \eq{eq:dA1d}, then for $Z(t=0)<0$ 
 the solution develops a finite-time singularity i.e. $Z\to -\infty$ in finite time $t=t_\star$.  
Such singularities have been named caustics~\cite{wilkinson2003path}.
We note that, in principle, \eq{eq:dA1d} is an inappropriate model for flows in clouds because 
the incompressibility constraint ($\dive\uu=0$) dictates $A$ to be identically zero. 
Let us, nevertheless, model the effects of turbulence in \eq{eq:dA1d} by
replacing $A$ by a Gaussian, white-in-time noise. 
This turns \eq{eq:dA1d} into a stochastic differential equation.  
From the  corresponding Fokker-Planck equation
Derevyanko et al. \cite{derevyanko2007lagrangian} evaluated the 
rate of formation of caustics
\begin{equation}
  J \sim \exp\left(- \frac{C}{\taup}\right) \/,
  \label{eq:J1d}
\end{equation}
where $C$ is a constant.
In two and three dimensions appearance of singularities implies that the
trace of the matrix $\mZ$ becomes infinitely
large~\cite{gus+meh16}. 
Blow-up of $\mZ$ implies that two nearby particles can have very high
relative velocity~\cite{gustavsson2011distribution,gustavsson2013distribution}.
In other words,  inertial particles can detach from the flow.
This effect is some times referred as sling effect~\cite{falkovich2002acceleration}.

\section{Summary of earlier works}
There is, by now, a significant volume of evidence
from direct numerical simulations of turbulence that shows that
a theory based on caustics correctly predicts the clustering and relative
velocities of heavy particles~\cite{
vosskuhle2014prevalence, per+jon15, bhatnagar2018statistics, bhatnagar2018relative,
 rani2019clustering, dhariwal2018small}.
Analytical calculations~\cite{falkovich2002acceleration,wilkinson2006caustic,
  derevyanko2007lagrangian, gustavsson2013distribution} of the rate of
formation of caustics have been limited to one dimensional flows till very
recently~\cite{meibohm2020paths}.
In such calculations the turbulence is approximated by a smooth random (Gaussian)
flow with a length scale $\ell$, correlation time $T$ and velocity scale $u$.
The gradient of flow velocity is assumed to be Gaussian. 
In addition to the Stokes number, this introduces another dimensionless number,
the Kubo number: $\Ku \equiv u T/\ell$.
In the limit $\Ku\to 0$ and $\St\to \infty$, such that the product $\Ku^2\St$
remains constant, the rate of formation of caustics is shown to be~\cite{wilkinson2006caustic}:
$J \sim \exp(-C/\St)$.
This limit also corresponds to solving \eq{eq:dA1d} with a flow-gradient $A$ that is
random, Gaussian and white-in-time~\cite{derevyanko2007lagrangian}. 
In the other limit -- correlation time of $A$ exceeds $1/A$ --
$J \sim \exp(-C/\St^2)$~\cite{falkovich2002acceleration}.
At finite Kubo numbers, Ref.~\cite{gustavsson2013distribution} used a model where $Z$
satisfies \eq{eq:dA1d} and $A$ satisfies an Ornstein--Uhlenbeck process. 
Numerical solution of this model and a WKB calculation showed that
that the rate of formation of caustics depends on both $\St$ and
$\Ku$: 
(a) At small Kubo numbers $J \sim \exp(-C/\St)$ for $St > 1$.
(b) at larger $\Ku$, e.g., at  $\Ku = 1$,
$J \sim \exp(-C/\St^2)$ in the limit $\St\to 0$. 

Can we apply the results from these simple models to understand
turbulent flows?
Note that incompressible turbulent flows are either two 
or three dimensional.
Numerical calculation for \eq{eq:dA} in two and three
dimensions~\cite{wilkinson2007unmixing} with
a white-in-time $\mA$ gives $J \sim \exp(-C/\St)$.
But turbulent flows are neither white-in-time nor
do then remain correlated up to large times. 
Then, one possible way to relate these calculations to turbulent flows
is to associate the characteristic length, velocity and the
time scale of these synthetic flows with the length, velocity,
and time scale at the Kolmogorov scale
(also called the dissipative scale) respectively; hence $\Ku = 1$. 
But even at the dissipative scales a turbulent flows cannot be described
by just one length scale~\cite{schumacher2007asymptotic}, in other
words there is not one unique Kolmogorov length scale.
Furthermore, in turbulent flows  the flow-gradient tensor, $\mA$,
obeys non-Gaussian statistics~\cite{chevillard2006lagrangian}.
Hence, in summary, it is not obvious how to relate these
analytical results to the turbulent flows. 
Cautiously, we expect the results in turbulent flows to be similar to those
obtained for $\Ku = 1$, i.e.,
$J \sim \exp(-C/\St^2)$ as $\St\to 0$.
As far as we are aware there has been so far no direct calculation of
rate of formation of caustics in two-dimensional turbulence, although
signatures of caustics has been observed in Eulerian simulations~\cite{mitra2018topology}. 
To the best of our knowledge, there is only one~\cite{falkovich2007sling}
calculation of the rate of formation of caustics in direct numerical
simulations of three-dimensional turbulence.
The calculation is for Taylor micro-scale Reynolds number of ranging from
about $45$ to $100$. 
The authors claim $J \sim \exp(-C/\St)$ but the conclusion is based
on fitting a nonlinear function with four parameters to a data with
seven points, see e.g. Ref.~\cite{dyson2004meeting} for a criticism of a similar exercise.
Here, we repeat their calculation for higher Reynolds numbers
and improved statistics,
in both two and three-dimension.
In three dimension we use $10$ million particles for each Stokes number. 

\section{Direct numerical simulations}

We solve the Navier--Stokes equation in two and three dimensions with heavy 
inertial particles, \eq{eq:HIP} in the flow.
In three dimensions we use the pencil-code~\cite{pencil-code,collaboration2021pencil}.
The same code has been used in several earlier
publications~\cite{bhatnagar2018statistics, bhatnagar2018relative}.
For the two-dimensional simulations we use a spectral code
which has also been used in several earlier
publications~\cite{pandey2019clustering}.
As the heavy inertial particles are much smaller than the energy containing
scales of the flow, we use the Kolmogorov length scale $\leta$ and
and time scale $\teta$ as our characteristic length and time scales respectively. 
We list the relevant parameters of our simulations in table~\ref{tab:runs}.
In clouds the Stokes number~\footnote{The Kolmogorov time scale in clouds
  in estimated based on estimates of the energy-dissipation-rate,  $\varepsilon$,
  which varies over a large range in different types of clouds.
  Consequently,  the range of Stokes number for particles of the  same size may be different
   in different clouds. The numbers we quote are typical of cumulus clouds.} 
  ranges from $0.01$ to $2$ for droplets
  of size $10$ to $60$ micrometer~\cite{sha03, ayala2008effects,gra+wan13}.  
We use $\St = 0.1$ to $3.1$ in three dimensions and
$\St = 0.12$ to $1.1$ in two-dimensions.
In Table~\ref{tab:stokes} we give a complete list of Stokes numbers used in each of
our simulations. 
In two dimensions, depending on which length-scale is being forced,
the turbulence may be dominated by either direct cascade of
enstrophy (e.g., run $\tt{2d-R1}$) or inverse cascade of
energy (e.g., run $\tt{2d-R8}$). 
In some two-dimensional simulations we have used a deterministic, 
Kolmogorov force in others we have used a stochastic, 
white-in-time, force. 
We have changed our Reynolds number over a large range,
from approximately $150$ to about $4000$.

In addition to tracking the heavy inertial particles, we solve
\eq{eq:dA} on each of the particles. 
We choose $\mZ=0$, the zero matrix, on all the particles at $t=0$.
We count how many times the trace of $\mZ$ ($\Tr{\mZ}$) crosses a large
negative threshold, $\Zth$ 
-- we have used several values for this threshold. 
We define, the sum of all such crossing events over all the particles up to
time $t$  to be $N(t)$.
The rate of formation of caustics 
\begin{equation}
J(t) = \lim_{t\to\infty}\frac{N(t)}{t} \/. 
\label{eq:Jt}
\end{equation}
\section{Results}
\begin{figure}
\includegraphics[width=0.48\linewidth]{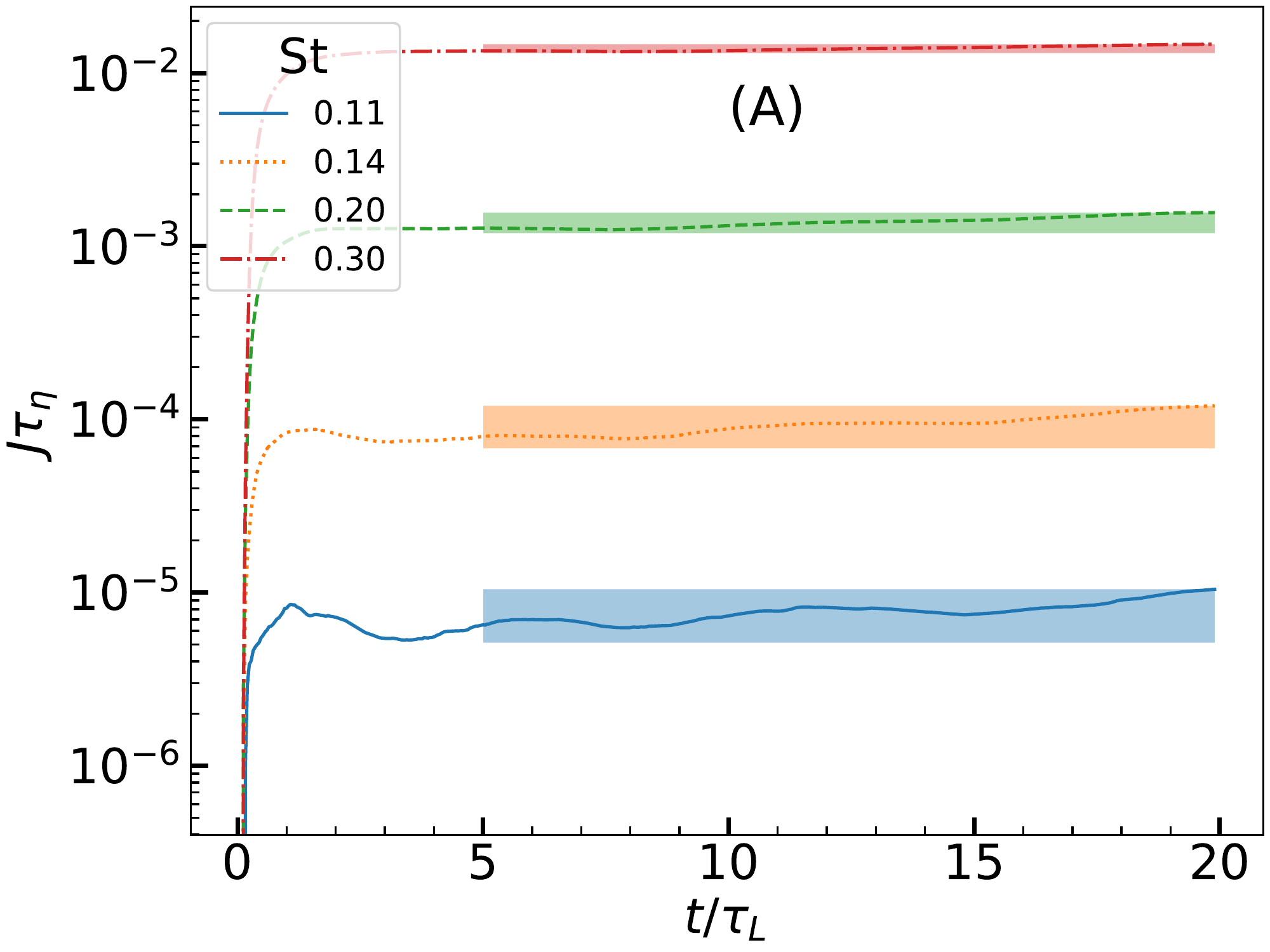}
\includegraphics[width=0.48\linewidth]{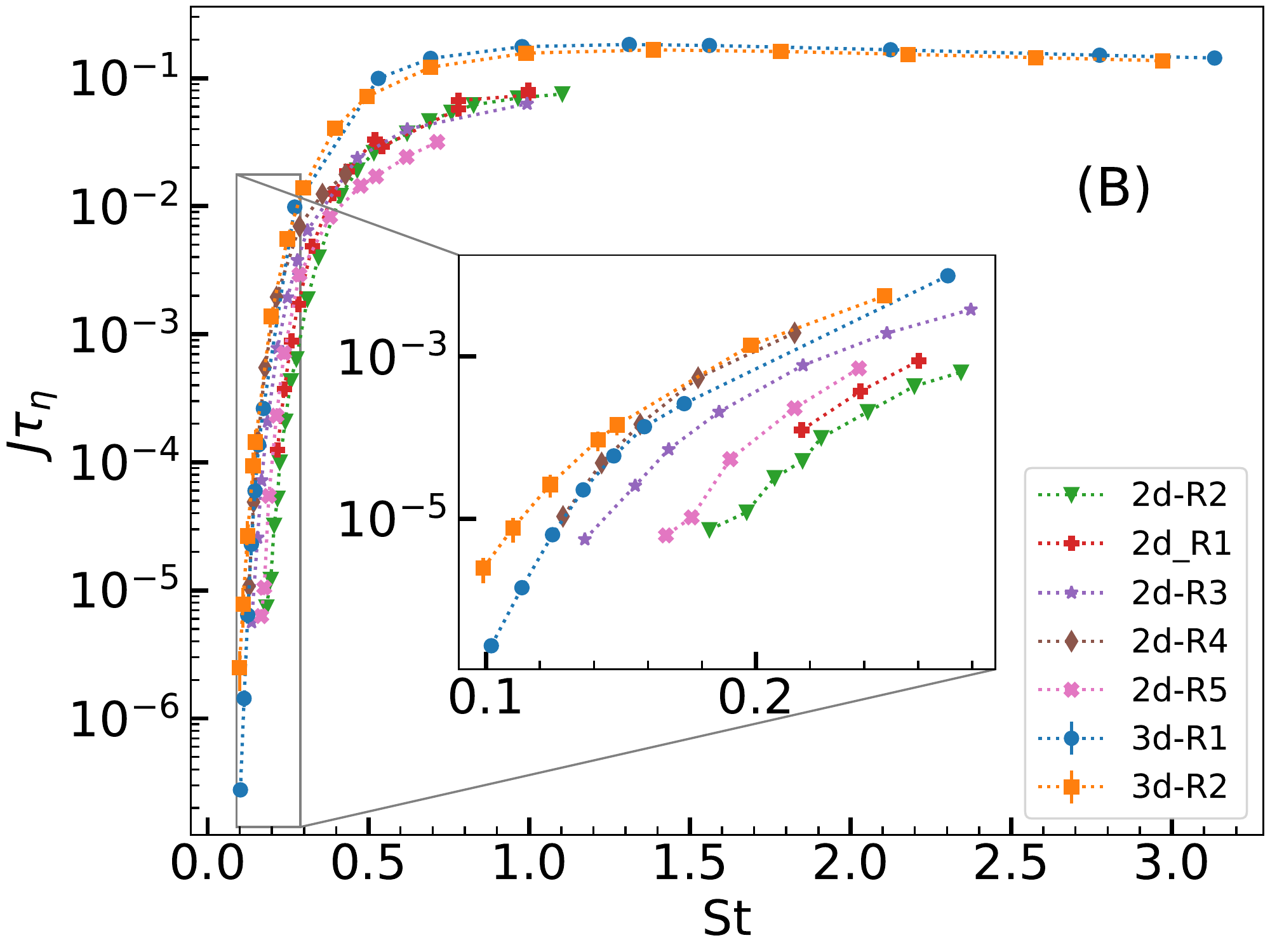}\\
\includegraphics[width=0.48\linewidth]{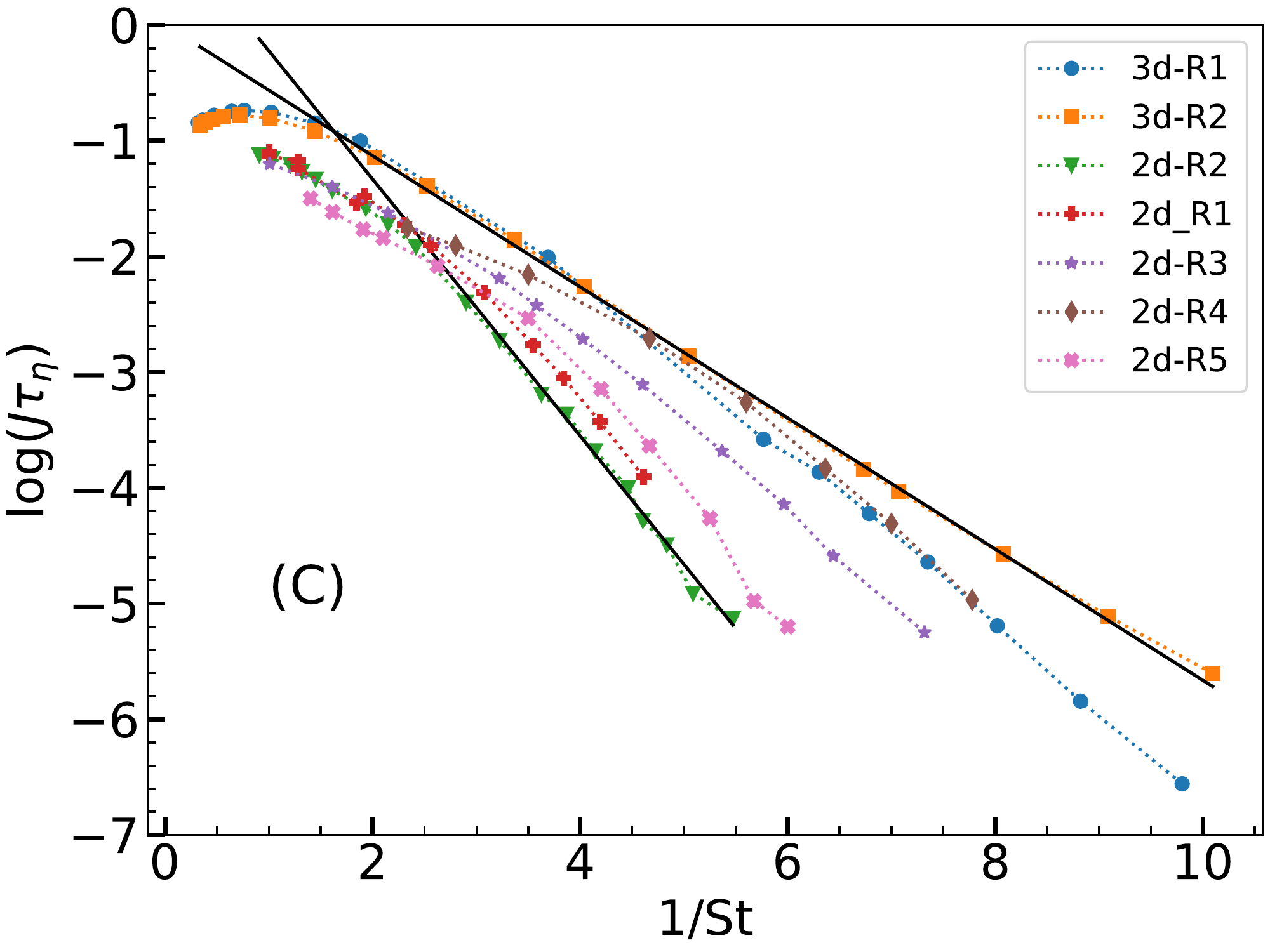}
\includegraphics[width=0.48\linewidth]{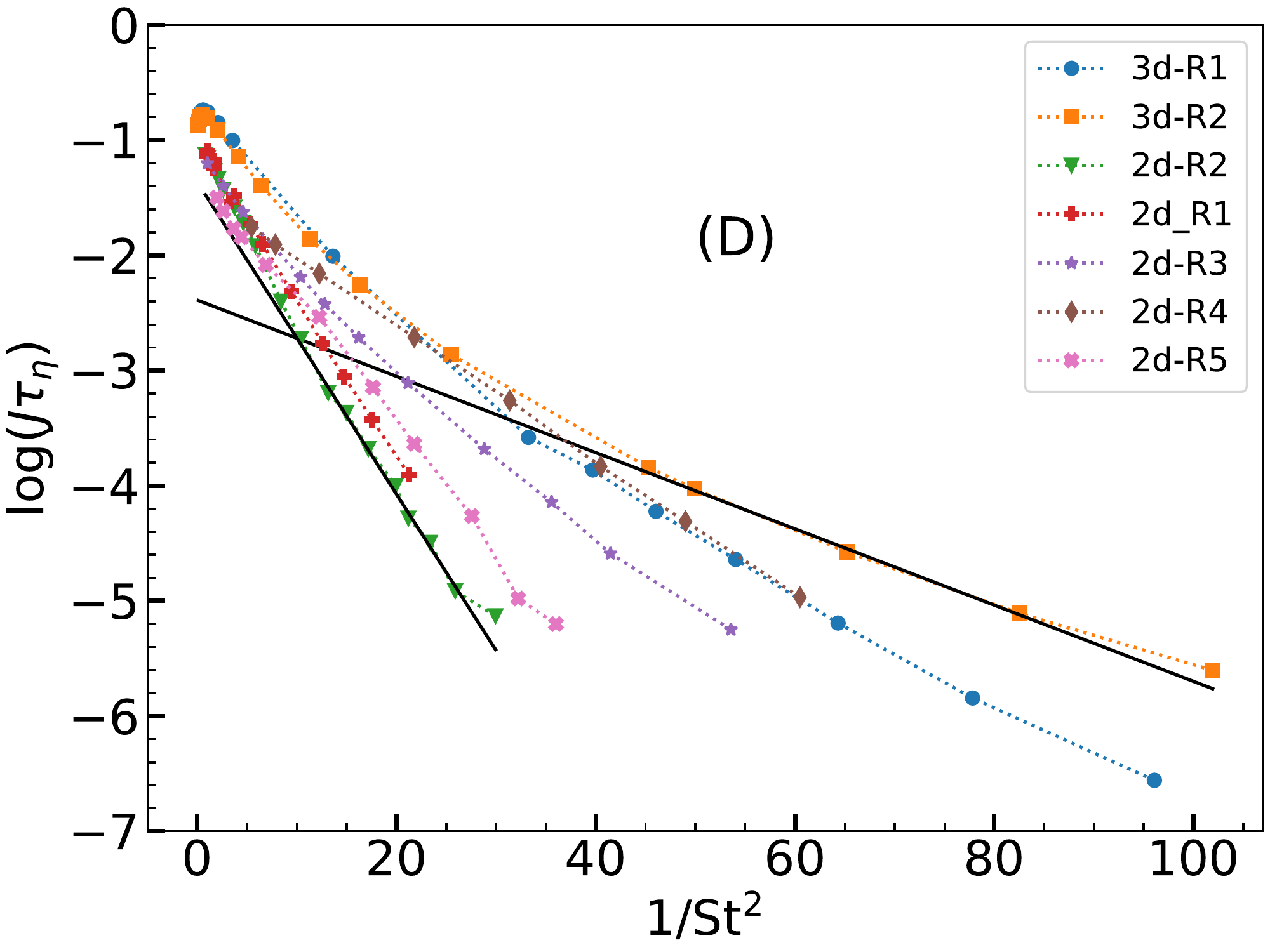}
\caption{\label{fig:JJ}(A) The rate of formation of caustics $J(t)$ versus $t$ for 
four different values of $\St$. We calculate the mean value of $J$ over the shaded region
in each time series and set the maximum and minimum value of $J$ as the
error bars. (B) The rate of formation of caustics $J(t)$ as a function of 
$\St$. The error bars are calculated according as shown in (A). They are
smaller than the symbols.
(C) The data in (B) plotted with $\log(\tau_{\eta}J)$ versus $1/\St$.
(D) The data in (B) plotted with $\log(\tau_{\eta}J)$ versus $1/\St^2$.
Notice that it is possible to find a range over which the plots show a linear trend
in both (C) and (D).
For clarity, we show such linear fits for two cases in each of (C) and (D). }
\end{figure}

In \subfig{fig:JJ}{A} we plot $J(t)$ versus $t$ for four
different representative values of the Stokes number.
At late times, $J(t)$ reaches a statistically stationary
state.
We consider the mean over this statistically stationary
state as our measurement of the rate of formation of
caustics and we use the maximum and minimum value of $J$
over this statistically stationary state to set error limits.
In \subfig{fig:JJ}{B} we plot the $J$ as a function of $\St$
for both two and three dimensional simulations.
In \subfig{fig:JJ}{C} we plot the logarithm of rate of formation of
caustics $\log(J)$ as a function of $1/\St$ for all the runs.
If \eq{eq:J1d} holds we expect to see a straight line as $1/\St$
becomes large.
Clearly it is possible to fit such a line to the data for both two-dimensional
and three-dimensional simulations.
However, the fit is better for the three dimensional cases. 
For clarity we show two such fits to our data in \subfig{fig:JJ}{D}.
But we do not consider this a conclusive evidence in support of \eq{eq:J1d}.
In \subfig{fig:JJ}{C} we change the abscissa to $1/\St^2$.
Here too it is possible to identify a region over which a straight line can be fit.
This time the fit is marginally better for the two dimensional cases. 
To conclude, our data do not unequivocally support either $J\sim \exp(-C/\St)$
or $J\sim \exp(-C/\St^2)$. 
\begin{figure}
\includegraphics[width=0.48\linewidth]{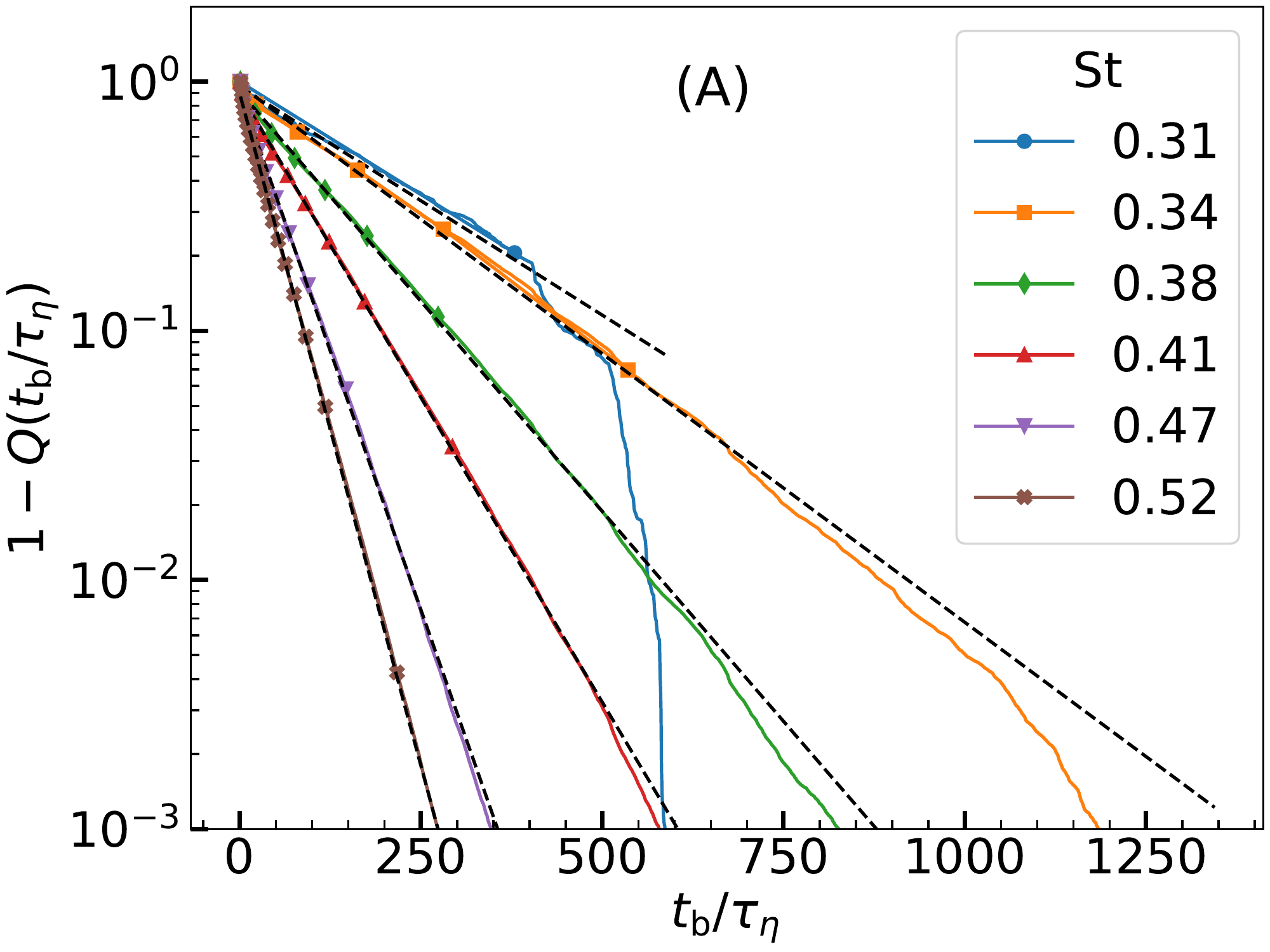}
\includegraphics[width=0.48\linewidth]{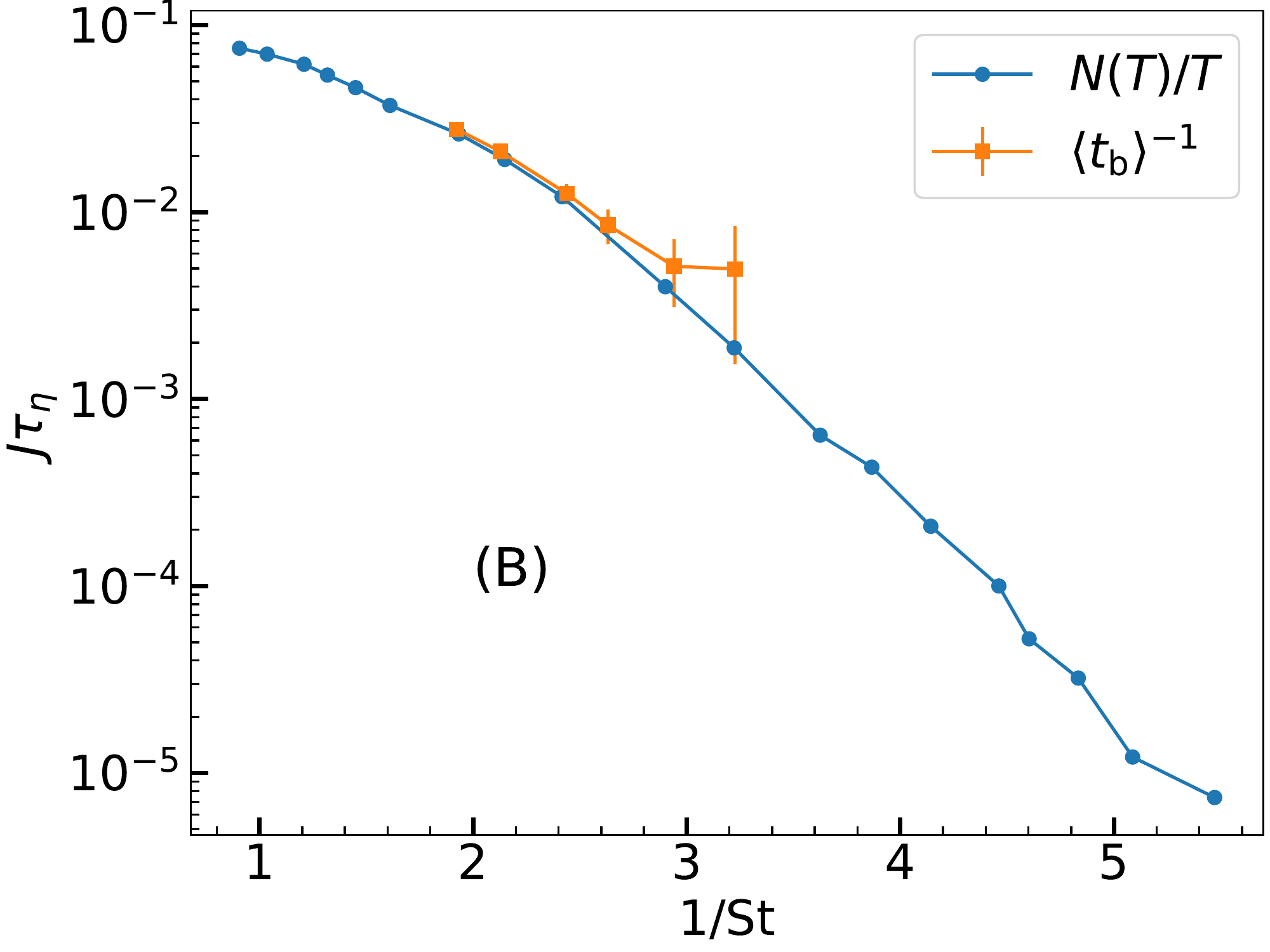}\\
\caption{\label{fig:cpdf}
(A) The cumulative probability distribution function, $Q(\tb)$
(calculated using the rank-order method) of the blow-up time $\tb$
for different values of the Stokes  number (Run {\tt 2d-R2}). 
The dashed lines are exponential fits to the tail of the data. 
The inverse of the mean of these data are plotted in (B) as a function
of $\St$ and compared with the rate $J$ calculated by counting the number of 
blow-ups, \Eq{eq:Jt}. } 
\end{figure}

Equation \ref{eq:Jt} is not the only way to define the rate-of-formation
of caustics. Let us  define the time it takes for the trace of $\mZ$
to exceed a threshold as the blow-up time $t\tb$.
An alternative definition of rate-of-formation of caustics is $1/\bra{\tb}$. 
Note that it is computationally more cumbersome to calculate the rate using
this later definition. 
We have done it for only one of our runs, $\tt{2d-R2}$. 
In \subfig{fig:cpdf}{A} we show the the cumulative probability distribution
of the blow-up time $Q(\tb)$ 
calculated using the rank-order method~\cite{sharma2017energization}, 
for several values of $\St$. 
Clearly the distribution has exponential tail. 
In \subfig{fig:cpdf}{B} we compare the two definitions of the rate-of-formation
of caustics -- they agree within error bars.

Singularities found in a numerical simulations are necessarily not true singularities
-- their detection depends on the threshold value we use. We have checked that
by changing our threshold value $-\Zth$ from $5$ to $10^3$ for the
three-dimensional runs and from $10$ to $10^{10}$ for the two-dimensional runs. 
The rate-of-formation of singularities itself changed, by small amounts, but its
dependence on Stokes remains essentially unchanged.
This is expected, because in \eq{eq:dA1d} once $\Tr{\mZ} < -1/\taup$ the dynamics
is determined by  $Z^2$. 
Hence any stochastic trajectories
of ${\mZ}$ where $\Tr{\mZ}$ becomes smaller than $-1/\taup$ will reach blowup.

The numerical calculation of the rate of formation of caustics is a
very difficult problem. 
We expect an asymptotic behaviour in the limit of small $\St$ but
in this limit the rate of formation of caustics become exponentially small. 
This implies that to obtain reliable statistics at small $\St$ we either
have to run our simulations for a very long time or for a very large number
of particles. 
We have done the largest direct numerical simulations for calculation of caustics so
far. We go up to $\Rey_{\lambda} = 180$ in three dimensions and
$\Rey_{\lambda} = 4100$ in two dimensions.
In three-dimensional simulations we have used $10$ million particles for each value
of Stokes.
In spite of such massive simulations we are unable to reach a definite conclusion
on the asymptotic behaviour of the rate of formation of caustics.
A recent paper~\cite{meibohm2020paths}, using analytical calculations
in a two-dimensional model, where the gradient of flow-velocity is
modeled by an Ornstein-Uhlenbeck  process,
shows that caustics tend to form for those particle trajectories that
experience low values of vorticity of the flow and large
values of rate-of-strain exceeding a threshold.
Furthermore,  two-dimensional direct numerical simulations agrees well
with this result.
In this paper, we have used the data from the same simulations but
have not been able to reach a definite conclusion about the rate-of-formation
of caustics.

Calculation of asymptotic behaviour from direct numerical simulations
of turbulence is generally a quite  challenging problem. 
For example, even in the case of stochastically forced Burgers equation in
one dimension, where it is possible to run very high resolution simulations
for a very long time, artifacts from sub-leading terms may
make simple biscaling masquerade as multiscaling~\cite{mit+bec+pan+fri05}.
A scaling regime over a scaling range of at least a decade is
considered necessary but not necessarily sufficient.
An alternative is to use a seminumerical procedure called
asymptotic extrapolation~\cite{pauls2007borel,van2009asymptotic,chakraborty2012nelkin}
that applies to the
data a sequence of suitable chosen transformation that
successively strip off dominant and subdominant terms.
Following Ref.~\cite{pauls2007borel}, we give a very short introduction
to this procedure here.

Consider the general problem where we investigate whether a given numerical
data can be fit with a function $G(r)$ with the leading order form
\begin{equation}
  G(r) \sim C r^{-\alpha}e^{-\delta r}\/,
    \label{eq:Gr}
\end{equation}
and thereby determine the parameters $C$, $\alpha$ and $\delta$.
One way is to ignore subleading corrections and try a least square fit.
This is what we have done so far to the function $J(\St)$.
The parameters we obtain in this fashion sometime depends crucially on the
fitting interval and determination of subleading corrections is almost
impossible.
The asymptotic extrapolation method allows, in principle, the determination
of the asympotic expansion of the function $G(r)$, including the
subleading terms, and an accurate determination of the parameters
provided the numerical data is obtained with very high precision. 
A crucial  feature of this procedure is that it uses the determination of
subleading terms to improve the accuracy on leading-order terms

\begin{figure}
  \includegraphics[width=0.48\linewidth]{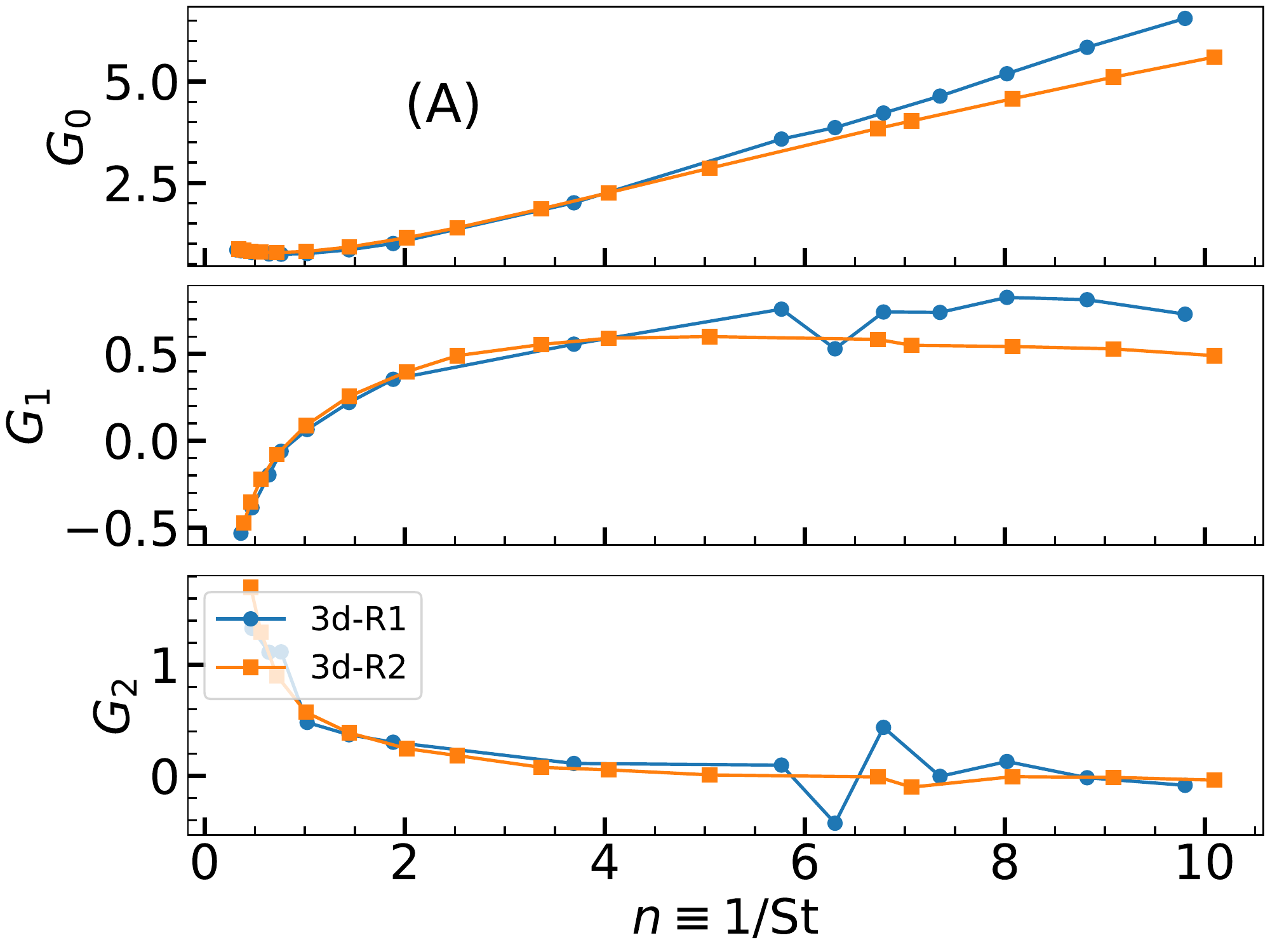}
  \includegraphics[width=0.48\linewidth]{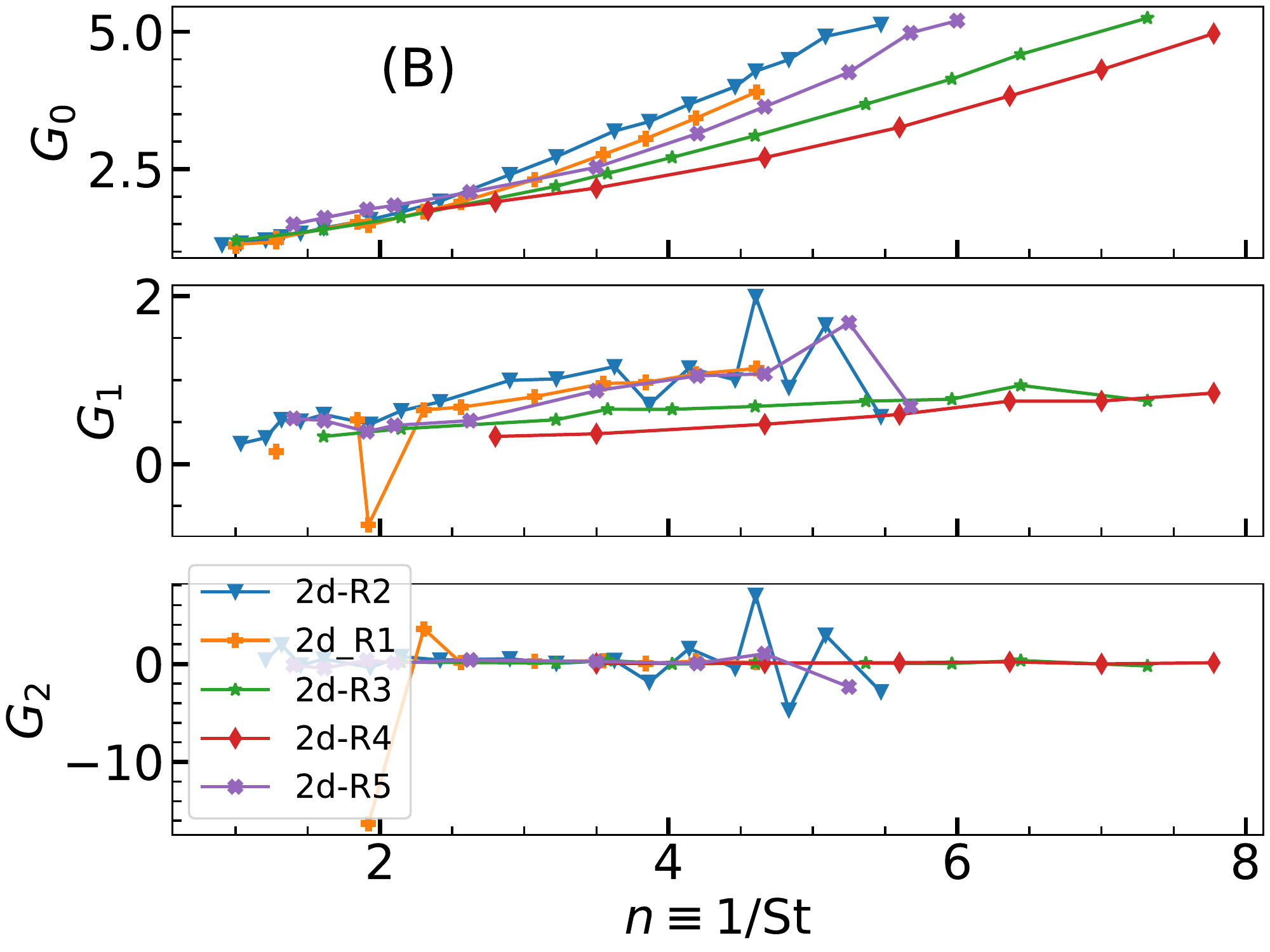}
  \caption{\label{fig:PF} Results of successive application of the
    asymptotic extrapolation on three dimensional (left) and two dimensional (right)
    data. The zeroth function $G_0(n) \equiv -\log(\teta J)$ and
    $n = 1/\St$. In the next iteration
    $G_1(n)  = D(G_0(n)) \equiv G_1(n) - G_1(n-1)$.
    In the next iteration
    $G_2(n)  = D(G_1(n))$. In both cases, $G_2$ is asymptotically zero although
    there is a significant noise in some cases. }
\end{figure}

We describe below how we apply this procedure to our data.
Let us assume that $J = \exp\left[-G(1/\St) \right]$
with
\begin{equation}
  G\left( n \right) \sim C n^{\alpha}(1 + \gamma_1 n + \gamma_2 n^2 \ldots )
\label{eq:fn}
\end{equation}
We intend to extract the coefficients $C$, $\alpha$, $\gamma_1$, etc
systematically from the numerical data $G(n)$.
We repeatedly iterate over $G$ with a discrete operator $\bm{D}$ such that
$G_{{\rm k+1}}(n) = \bm{D} G_{\rm k}(n) \equiv  G_{\rm k}(n) - G_{\rm k}(n-1)$
and $G_{\rm 0}(n) = G(n)$. 
We show the results of repeated application of this operator on our data
in \fig{fig:PF}.
The left panel shows the results for our three dimensional simulations and
the right panel shows the results for our two dimensional simulations.
In both cases, we notice that $G_{\rm 2}$ is practically zero although in some
cases with large noise.
Inverting the operators imply that $G(n) \sim C n $.
Any coefficient of higher power of $n$ is not discernible from the data.
Consequently, our data, in both two and three dimensions,
support $J \sim \exp(-C/\St)$.

A word of caution is necessary here. The technique of asymptotic extrapolation
is typically applied to data with high precision, otherwise repeated
application of discrete operators, e.g., the difference operator we used,
can get rid of the significant digits in the data and leave only noise.
It has been used to study scaling of moments of gradients in
direct numerical simulations of 
one dimensional Burgers equation~\cite{chakraborty2012nelkin}
with both double and quadruple precision.
For the case of double precision only three stages of iterations were possible.
We have used double precision in all our calculations.
We can iterate twice before the data became practically zero. 
This does not imply that there is no subdominant factor to $G(1/\St)$.
It rather means that the quality of our data does not allow us to
extract any such contribution. 

\section{Conclusion}
We have performed the highest resolution and most detailed study so far of
the  rate of formation of caustics in two and three-dimensional simulations
of heavy inertial particles in turbulence.
In spite of our diligence we have been unable to uncover how the rate of
formation of caustics depends on the Stokes and Reynolds number.
We find that a least-square fit to the data does not support
unequivocally either of the two possibilities
$J \sim \exp(-C/\St)$ or $J\sim \exp(-C/\St^2)$.
Next we applied the technique of asymptotic extrapolation to our data
to find $J \sim \exp(-C/\St)$.
We do not consider this conclusive evidence.
It is necessary to perform simulations at even smaller $\St$
with higher precision (at least quadruple precision).
In the absence of definitive numerical evidence this remains an open problem. 

  \begin{table*}[!ht]
  \caption{\label{tab:runs} Parameters used in both our three dimensional simulations
      are listed here. The two-dimensional runs are marked with the prefix $\tt{2d}$ and the
      three dimensional ones are marked with the prefix $\tt{3d}$.
      For the two dimensional 
      runs we use a variety of different situations. Some of them are forced with a
      deterministic Kolmogorov force, some are forced with a stochastic force.
      The force is always limited to a a single wave-number $\kf$. For a small $\kf$
      we develop a direct cascade, whereas for a large $\kf$ we develop a large
      range of inverse cascade. We also change the
      threshold of detection of singularities, $\Zth$, which has no appreciable effect on the
      dependence of $J$ (rate of formation of singularities) on the Stokes number.
      The rate of formation of caustics from these runs are plotted in \fig{fig:JJ}. 
      Definition of symbols:
      $\tt{N}$,  number of grid points in one direction, 2-d runs have $\tt{N}^2$ number
      of grid points and 3-d runs have $\tt{N}^3$ grid points;
      $dt$, time-step used in the 2-d solver. 
      $\nu$, viscosity;
      $\mu$, Ekman friction;
      $\fzero$, amplitude of the force;
      $\bra{\cdot}$, spatial average over the computational box and temporal
      average over the statistically-stationary, non-equilibrium, state of turbulence; 
      $\urms = \sqrt{\bra{\bm{u}^2}/d}$ ($d=2,3$ is the dimension),
      the root-mean-square velocity of the flow;
      $\kf$, the forcing wavenumber;
     $\LI \equiv 2\pi/\kf$, the integral scale;
     $\tL = \LI/u_{\rm rms}$, the large-eddy-turnover-time;
     $\bm{\omega} = \curl \uu$, the vorticity; 
     $\Omega = \bra{\bm{\omega}^2}/2$, the enstrophy;
     $\orms = \sqrt{(2/3)\Omega}$ in 3-d and  $\orms = \sqrt{2\Omega}$ in 2-d
      is the root-mean-square vorticity; 
      $\varepsilon = 2 \nu \Omega$, the rate of energy dissipation;
      $\eta \equiv (\nu^3/\varepsilon)^{1/4}$ , Kolmogorov (dissipation) length scale,
      $\lambda =\urms/\orms$, the Taylor microscale;
      $\Rel = \urms \lambda/\nu$, the Taylor microscale Reynolds number;
      $\St=\taup/\teta$, the Stokes number.
      We use $\St = 0.1$ to $3.1$ in three-dimensions and $\St = 0.12$ to $1.1$ in two
      dimensions. All the values of $\St$ used in different runs are given in
      table~\ref{tab:stokes}. 
      }

  	\begin{tabular}{ccccccccccccc}  
  	$runs$       &  ${\tt N}$ &  $\Rel$ & $\teta$ & $\eta$& $ \kf$  & $\tL$ &$\mu$ & $dt$ \\ \hline\hline

  			
  			
  			
  	$\tt{2d-R1}$ & 512  & 395  &  4.6   & 0.007  & 4 & 36.8 & 0.01 & 5$\times10^{-3}$ \\
  	
  	$\tt{2d-R2}$ & 1024 &  1311 & 2.9 & 0.005     & 4 & 21.0 & 0.01 & 5$\times10^{-3}$ \\
  			
  			  			
  	$\tt{2d-R3}$ & 1024 &  360 &  1.6  & 0.004  & 35  & 3.7 & 0.001 & 5$\times10^{-3}$ \\
  			
  	$\tt{2d-R4}$ & 1024 &  73 &  1.4  & 0.0037 & 100  & 1.0 & 0.0002 & 5$\times10^{-3}$ \\ 
  			
  	$\tt{2d-R5}$ & 1024 &  4100 &  2.1  & 0.005  & 2 - 3  & 17.7 & 0.01 & 1$\times10^{-3}$ \\ 
  			
  			
  			
  	$\tt{3d-R1}$ & 512 &  90    &  0.39  & 0.014 & 5  & 5.46 & --- & --- \\

  	$\tt{3d-R2}$ & 512 &  170   &  1.56  & 0.014 & 2  & 36.98 & --- & --- \\
  	\end{tabular}
  \end{table*}

  \begin{table*}[!ht]
  	\caption{\label{tab:stokes} Values of $\St$ used in different
		simulations
        }
\begin{tabular}{c|c} 
  \hline\hline 
  $runs$   &  $\St$ \\ \hline\hline
  $\tt{2d-R1}$  & 0.22, 0.24, 0.26, 0.28, 0.32, 0.39, 0.43, 0.52, 0.54, 0.78 \\
  $\tt{2d-R2}$  & 0.18, 0.20, 0.21, 0.22, 0.24, 0.26, 0.28, 0.31, 0.34,
                  0.41, 0.46, 0.52, 0.62, 0.69, 0.76, 0.83, 0.97, 1.10\\
  $\tt{2d-R3}$  & 0.14 0.15,  0.17,  0.19, 0.22, 0.25, 0.28, 0.31, 0.47, 0.62, 1.00 \\
  $\tt{2d-R4}$  & 0.13, 0.14, 0.16, 0.18, 0.21, 0.29, 0.36, 0.43  \\
  $\tt{2d-R5}$  & 0.17, 0.18, 0.19, 0.21, 0.24, 0.29, 0.38, 0.48, 0.52, 0.62, 0.71 \\
  $\tt{3d-R1}$  & 0.10, 0.11, 0.12, 0.14, 0.15, 0.16, 0.17,
                  0.27, 0.53, 0.69, 0.98, 1.31, 1.56,2.12, 2.77, 3.13\\
  $\tt{3d-R2}$  & 0.10, 0.11, 0.12, 0.14, 0.15, 0.20, 0.25, 0.30,
                  0.40, 0.50, 0.70, 0.90, 1.40, 1.79, 2.19, 2.59\\
  \hline\hline 
\end{tabular}
\end{table*}
  \enlargethispage{20pt}


\aucontribute{VP and AB have contributed equally to this work.
  DM and AB wrote the code for simulations in three dimensions. VP and PP wrote the code for
  simulations in two dimensions. All the four authors analyzed the data.
  DM drafted the manuscript with inputs from all the other authors.
  All authors read and approved the manuscript.}
\competing{The authors declare that they have no competing interests.}
\funding{ PP and VP acknowledge support from
  DST (India) Project Nos. ECR/2018/001135 and  DST/NSM/R\&D\_HPC\_Applications/2021/29. 
The computations were enabled by resources provided by the Swedish National Infrastructure for 
Computing (SNIC) at PDC partially funded by the Swedish Research Council through grant 
agreement no. 2018-05973.
AB and DM acknowledges financial support from the grant Bottlenecks for 
particle growth in turbulent aerosols from the Knut and Alice Wallenberg 
Foundation (Dnr. KAW 2014.0048) and from Swedish Research Council 
Grant no. 638-2013-9243 as well as 2016-05225. }
\ack{We have used matplotlib~\cite{Hunter:2007} to generate the figures in this paper.
We thank Jan Meibohm and Bernhard Mehlig for discussions.} 
\bibliographystyle{rsta}

\end{document}